\documentclass[aps,prb,twocolumn,groupedaddress,showpacs]{revtex4}

\usepackage{dcolumn}
\usepackage{graphicx,amssymb,amsmath}
\usepackage{bm}
\usepackage{natbib}
\usepackage{epstopdf}
\begin{document}

\title{Exact longitudinal plasmon dispersion relations for one and two dimensional Wigner crystals}

\author{Shimul Akhanjee}
\email[]{shimul@physics.ucla.edu}
\affiliation{Department of Physics, UCLA, Box 951547, Los Angeles, CA 90095-1547}


\date{\today}

\begin{abstract}
We derive the exact longitudinal plasmon dispersion relations, $\omega(k)$ of classical one and two dimensional Wigner crystals at $T=0$ from the real space equations of motion, of which properly accounts for the 
full unscreened Coulomb interactions. We make use of the polylogarithm function
in order to evaluate the infinite lattice sums of the electrostatic force constants. From our exact results
we recover the correct long-wavelength behavior of previous approximate methods. 
In 1D, $\omega(k) \sim \left| k \right|\log ^{1/2} (1/k)$, validating the known RPA and bosonization form. In
2D $\omega(k) \sim \sqrt k$, agreeing remarkably with the celebrated Ewald summation result. Additionally,
we extend this analysis to calculate the band structure of tight-binding models of non-interacting electrons with arbitrary power law hopping. 

\end{abstract}

\maketitle

\section{Introduction}

In the realm of one-dimensional(1D) physics, many analytical solutions for the interacting
electron gas exist at various regimes of electron-electron interaction strength, both repulsive and attractive, often revealing non-trivial fermionic 
instabilities\cite{quantum}. 
Wigner crystals are one such phase, existing in the extreme limit of electronic correlations\cite{wigner}. 
Wigner crystallization is also known to occur even in classical systems in higher dimensions for sufficiently strong unscreened
Coulomb repulsion, however for the specific case of 1D electrons, true long-ranged order is not possible even with longer ranged interactions.

Unlike Fermi liquid systems, for Wigner crystals, the precise mathematical and experimental behavior of the elementary excitations are still
far from completely understood. The charged collective modes at $T=0$ are one such entity, where its essential features are captured by a classical 
description\cite{giampwc}. The spin degrees of freedom also lead to another type of collective mode, of which is important in describing
the magnetic properties. However, for simplicity we assume that the spin wave transport is sufficiently decoupled from the plasmon propagation.

The effects of long range interactions are often neglected in many discussions of Wigner crystals even though one would expect in real
physical systems a complete breakdown of screening. So far, the complete plasmon dispersion and more importantly, its precise long wavelength behavior
has not been deduced analytically.
We present a simple
derivation of the plasmon dispersion relation in a 1D Wigner crystal phase that exploits the summability properties of power law interactions that 
are unique to 1D systems. Morevover, the correct behavior of the two dimensional(2D) longitudinal eigenmode is also calculated exactly by performing  
the neccessary single summation, of which produces excellent agreement with known numerical results.
Additionally, as an example of the utility of this form of analysis and owing
to a similar mathematical structure
we also apply these methods to calculate the band structure of an electronic tight-binding model with power law interactions. This is shown in the appendix. 

\section{The Dispersion Relations}

Let us begin with a classical 1D array of $L$ particles interacting with unscreened, 
long-range Coloumb forces, described by the Hamiltonian: 

\begin{equation}
H = \sum\limits_{i = 1}^L {\frac{{p_i^2 }}{{2m_e^* }} + \frac{1}{2}} \sum\limits_{i \ne j}^{} {\frac{{e_0^2 }}{{\left| {x_i  - x_j } \right|}}}
\label{eq:plashamilt} 
\end{equation}
In the above equation we have defined $p_i$ and $m_e^*$ as the particle momentum and effective mass of the ith particle, respectively. Our system lives in the low density regime in which the Coulomb interactions are much larger than the kinetic energy resulting in the crystalline ordering of the 
particles with the spatial coordinates $\lbrace x_i \rbrace$ separated with lattice constant $a$. Moreover, the system is stabilized by a
positive Jellium neutralizing background. The eigenvalue equation of interest follows directly from the classical equations of motion for the total force
acting on each particle. This is given by the following expression:

\begin{equation}
-m_e \omega ^2 u(x) + \sum\limits_{x' \ne x}^{} {\phi _{x,x'} u(x')}  = 0
\label{eq:plaseigen}
\end{equation}
where $u(x)$ is the displacement of a lattice site from equilibrium and,

\begin{equation}
{{\phi _{x,x'}  = 1} \mathord{\left/
 {\vphantom {{\phi _{x,x'}  = 1} {\left( {x - x'} \right)^3 }}} \right.
 \kern-\nulldelimiterspace} {\left( {x - x'} \right)^3 }}
\label{eq:electro}
\end{equation}
is the electrostatic force constant between two particles in the array, of which is the second derivative of the
interaction potential. As a consequence of the periodic ordering and translational symmetry present in the system we can assume the eigenfunctions have the form,
\begin{equation}
u(na) \propto \exp[i(kna - \omega t)]
\label{eq:eigenfunc}
\end{equation}
of which $k$ is the Fourier component and $n = 1,2,3 ...$. We substitute equations (\ref{eq:electro}) and (\ref{eq:eigenfunc}) into (\ref{eq:plaseigen}),

\begin{equation}
\omega ^2 \propto \sum\limits_{r = 1}^\infty  {\frac{{\sin [kr/2]^2}}{{r^3 }}}  
\label{eq:subtract} 
\end{equation}
where we have defined $r \equiv na$ and have set $m_e = e_0 = 1$ for simplicity. Thus, it is our primary task to carry out the infinite summation of equation (\ref{eq:subtract}). Hitherto, the most common approaches have been approximate, using 
methods such as the Ewald summation technique\cite{ewald}, where a solution is presented in terms of rapidly converging sums. We make use of the Polylogarithm function $Li_n (z)$ also known as the de Jonquières function, defined as\cite{stegun}:

\begin{equation}
Li_n (z) = \sum\limits_{k = 1}^\infty  {\frac{{z^k }}{{k^n }}} 
\label{eq:polylog}
\end{equation}
This definition may be extended to all of the complex plane through analytic continuation, therefore we apply equation (\ref{eq:polylog}) to the summation of 
equation (\ref{eq:subtract}) yielding,

\begin{equation}
\begin{aligned}
\omega_{1D}(k) &\propto \sum\limits_{r = 1}^\infty  {\frac{{\sin [kr]^2 }}{{r^3 }}} \\
&= \frac{1}{2}\left( { - Li_3 (e^{ - ik} ) - Li_3 (e^{ik} ) + 2\zeta (3)} \right) \\
\end{aligned}
\label{eq:1ddisp}
\end{equation}
where  $\zeta(x)$ is the Reimann zeta function. At long wavelengths the polylogarithms can be expanded to the lowest order, yielding:

\begin{equation}
\omega_{1D}(k)  \approx \left| k \right|\log^{1/2} [1/\left| k \right|]
\label{eq:1dshort}
\end{equation}

Let us turn our attention to the two dimensional case, for which equation (\ref{eq:plaseigen}) can be generalized to a double summation over a tensor. The dominant contribution 
to the longitudinal eigenmode is the following sum:

\begin{equation}
\begin{aligned}
\omega _{2D} (k) &\propto \sum\limits_{r = 1}^\infty  {\frac{{\sin [kr]^2 }}{{r^2 }}}  \\
&= \frac{1}{{12}}\left( {\pi ^2  - 3Li_2 (e^{ - ik} ) - 3Li_2 (e^{ik} )} \right)\\
\end{aligned}
\end{equation}
We can further simplify this expression by making use of the following identity,
\begin{equation}
Cl_n (x) = \left\{ \begin{array}{l}
 \frac{1}{2}i[Li_n (e^{ - ix} ) - Li_n (e^{ix} )] \to n - even \\ 
 \frac{1}{2}[Li_n (e^{ - ix} ) + Li_n (e^{ix} )] \to n - odd \\ 
 \end{array} \right.
\label{eq:clausenid}
\end{equation}
where $Cl_n(x)$ are Clausen functions\cite{stegun} for a given $n$. 
It is known from functional analysis that certain Clausen functions have an exactly summable representation for arguments in a restricted range\cite{stegun}.
In particular for $0 \le k \le 2\pi$,

\begin{equation}
Cl_2 (k) = \frac{\pi^2 }{6} - \frac{{\pi k}}{2} + \frac{{k^2 }}{4}
\label{eq:lclausen}
\end{equation}
Apparently, the periodicity of our system guarantees that the values of $k$ are restricted to the first Brillouin zone. Therefore, a more convenient
representation of the 2D longitudinal plasma dispersion relation becomes:

\begin{equation}
\omega _{2D} (k) \propto \sqrt {\frac{{\pi \left| k \right|}}{2} - \frac{{k^2 }}{4}} 
\end{equation}

Apparently the long wavelength behavior reduces to:

\begin{equation}
\omega_{2D} ( \left| k \right| )  \propto \sqrt {\left| k \right|} 
\label{eq:2dshortdisp}
\end{equation}

If we place this derivation in the context of earlier work, this classical result can be compared to the attempts by other authors using a quantum-mechanical treatment of charged collective modes with long-range interactions.
Until now, no exact analytical results exist for the classical plasmon dispersion relations of Wigner crystals in any dimension. Gold and Ghazali\cite{gold} examined a
correlated quasi-1D electron system by using the Random Phase Approximation(RPA). In the RPA treatment the authors remedy the diverging Fourier transform of the $1/r$
potential in 1D by phenomenologically adding a small but finite system width $d$ that leads to a logarithmic part of the interaction, separating the short-ranged behavior from
the long-ranged one. The resulting charged modes have the following dispersion,

\begin{equation}
\omega _{RPA} (k) \approx \frac{{2e_0 }}{{\sqrt \pi  }}\sqrt {v_f } \left| k \right|\log ^{1/2} \left( {\frac{1}{{kd}}} \right)
\label{eq:rpadisp}
\end{equation}
where $v_f$ is the Fermi velocity. Clearly a notable difference between our classical result (\ref{eq:1ddisp}) and the RPA result(\ref{eq:rpadisp}) is the logarithmic
singularity in the limit $d \to 0$, of which is a direct consequence of the authors considering a quasi-1D system rather than the purely 1D system that we have just discussed. 
Although our exact result contains extra dispersive curvature at values of $k$ near the Brillouin zone boundary, 
our classical summation technique agrees with the RPA result's long-wavelength behavior.  

\begin{figure}
\centerline{\includegraphics[height=2.3in]{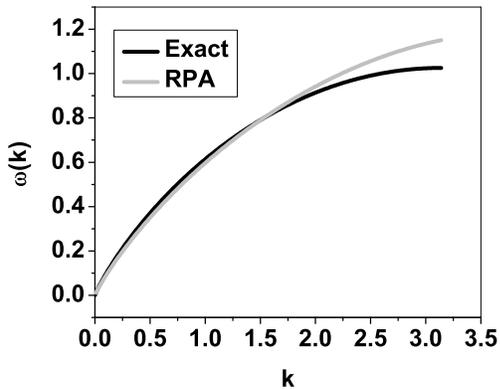}}
\caption{A comparison of the exact  1D classical dispersion with the RPA result.}
\label{fig:1d}
\end{figure}

\begin{figure}
\centerline{\includegraphics[height=2.3in]{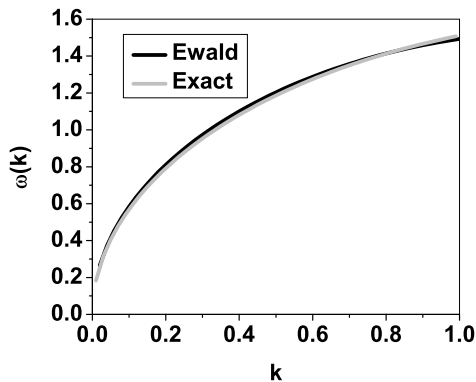}}
\caption{A comparison of the exact 2D longitudinal dispersion with the Ewald result.}
\label{fig:2d}
\end{figure}

Still, others have attempted to describe 1D WC behavior in the limit of an elastic Hamiltonian such as in the Luttinger liquid phase(Schulz, 1993)\cite{schulzwigner}. 
The Luttinger liquid arises as an instability in a 1D electron system with strong short ranged interactions, captured by the Hubbard Hamiltonian. Again 
they contend that the true long range behavior of the Coulomb interaction is not important aside from these minor logarithmic correction factors that modify the elastic modes to
produce a dispersion relation that has the same behavior as the RPA result(\ref{eq:rpadisp}). A comparison is show in Figure \ref{fig:1d}.

In 2D, until now there have been no analytical results for the precise behavior of the longitudinal plasma eigenmode. The Ewald technique has been 
numerically implemented by previous authors\cite{bonsall,crandall}. They discovered an unusual $\omega(k) \sim \sqrt k$ dependence arising strictly from the long-ranged interactions,
as it is known that short-ranged interactions produce a linear phonon dispersive form. We compare our closed form result with the Ewald summation technique
in Figure \ref{fig:2d}. Evidently, there is excellent agreement.

\section{Concluding Remarks}

We have derived the complete dispersion relation for a longitudinal plasmon in 1D and 2D Wigner crystals with unscreened Coulomb interactions. Our
analysis introduces mathematical methods for analytically evaluating a certain class of lattice sums that have traditionally been performed numerically. 
Recently, Wigner crystals in lower dimensions have experienced a 
resurgence in interest by both experimentalists and theorists alike \cite{wire}, in systems such as low density quantum wires and various soft condensed matter systems\cite{boris}.
Furthermore, the precise wavevector dependence of the eigenfrequencies have important consequences for many physically pertinent quantities such as the 
dynamical response functions, and thus the results of this paper provide some degree of analytical control in future investigations of Wigner Crystals that emphasize the role of long-ranged interactions.

\begin{acknowledgements}

I would like to thank J. Rudnick for his crucial assistance on this work and Y. Tserkovnyak for his comments on the manuscript.

\end{acknowledgements}

\begin{appendix}

\section{The band structure for power law hopping}

The tight-binding Hamiltonian of interest takes the following form,

\begin{equation}
H = \sum\limits_i^L {\varepsilon _0 \left| i \right\rangle \left\langle i \right|}  + \sum\limits_{\left\langle {ij} \right\rangle }^L {t_{ij} \left( {\left| i \right\rangle \left\langle j \right| + \left| j \right\rangle \left\langle i \right|} \right)} 
\label{eq:hamilt}
\end{equation}
where ${\left\langle {ij} \right\rangle }$ denotes a sum over all unique pairs of lattice sites $i$ and $j$. For simplicity we have assumed 
only one orbital per site. The hopping matrix elements $t_{ij}$ have the power law dependence,

\begin{equation}
t_{ij}  = \frac{{t_0 }}{{\left| {i - j} \right|^\beta  }}
\label{eq:hopping}
\end{equation}
Evidently, the Hamiltonian satisfies the time-independent Schrodinger equation $H\left| {\psi _k } \right\rangle  = E(k)\left| {\psi _k } \right\rangle$. 
Furthermore, for translationally invariant systems, 
we can make use of the Fourier transform, having the form $\left| {\psi _k } \right\rangle  = \sum\limits_{l'} {e^{ikl'} } \left| {l'} \right\rangle$, 
with $l = 1,2,\dots$. We can apply this ansatz along with equation (\ref{eq:hamilt}) to the Schrodinger equation to yield the following 
expression,

\begin{equation}
\begin{aligned}
 H\left| {\psi _k } \right\rangle  = \sum\limits_{l'} {e^{ikl'} } \left| {l'} \right\rangle \left\{ {\sum\limits_{r = 1}^\infty  {\frac{{t_0 (\cos [kr] - 1)}}{{r^\beta  }} + \varepsilon _0 } } \right\} \\ 
  = \sum\limits_{l'} {e^{ikl'} } \left| {l'} \right\rangle \left\{ {t_0 (Li_{\beta}[e^{ik} ] + Li_{\beta}[e^{ - ik} ]) + \varepsilon _0 } \right\} \\ 
 \end{aligned}
\end{equation}
where we have defined $ r = \left| i - j \right| $ and we have also absorbed various constants into the definition of $t_0$ and $\epsilon_0$.
Again we made use of the polylogarithm function $Li_n (z)$ and following similar steps as before
our final expression for the energy dispersion relation reduces to:

\begin{equation}
E^{\beta}(k) = t_0 (Li_{\beta}[e^{ik} ] + Li_{\beta}[e^{ - ik} ]) + \varepsilon _0 
\label{eq:tbfinal}
\end{equation}
Equation (\ref{eq:tbfinal}) is general to a particular exponent $\beta$, and one must properly consider the odd and even cases in order to determine
when it is appropriate use a particular Clausen function. More specifically, for odd powers of $\beta$ one may use the equation (\ref{eq:clausenid})
reducing the final energy dispersion to:

\begin{equation}
E^\beta  (k) = 2t_0 Cl_\beta  [k] + \varepsilon _0 
\end{equation}
for $\beta = 2n+1$; n=,0,1,2$\dots$.

\end{appendix}


\begin{thebibliography}{11}
\expandafter\ifx\csname natexlab\endcsname\relax\def\natexlab#1{#1}\fi
\expandafter\ifx\csname bibnamefont\endcsname\relax
  \def\bibnamefont#1{#1}\fi
\expandafter\ifx\csname bibfnamefont\endcsname\relax
  \def\bibfnamefont#1{#1}\fi
\expandafter\ifx\csname citenamefont\endcsname\relax
  \def\citenamefont#1{#1}\fi
\expandafter\ifx\csname url\endcsname\relax
  \def\url#1{\texttt{#1}}\fi
\expandafter\ifx\csname urlprefix\endcsname\relax\def\urlprefix{URL }\fi
\providecommand{\bibinfo}[2]{#2}
\providecommand{\eprint}[2][]{\url{#2}}

\bibitem[{\citenamefont{Giamarchi}(2003)}]{quantum}
\bibinfo{author}{\bibfnamefont{T.}~\bibnamefont{Giamarchi}},
  \emph{\bibinfo{title}{Quantum Physics in One Dimension}}
  (\bibinfo{publisher}{Clarendon Press}, \bibinfo{address}{Oxford},
  \bibinfo{year}{2003}).

\bibitem[{\citenamefont{Wigner}(1958)}]{wigner}
\bibinfo{author}{\bibfnamefont{E.}~\bibnamefont{Wigner}},
  \bibinfo{journal}{Phys. Rev.} \textbf{\bibinfo{volume}{46}},
  \bibinfo{pages}{1002} (\bibinfo{year}{1958}).

\bibitem[{\citenamefont{Chitra et~al.}(2002)\citenamefont{Chitra, Giamarchi,
  and Doussal}}]{giampwc}
\bibinfo{author}{\bibfnamefont{R.}~\bibnamefont{Chitra}},
  \bibinfo{author}{\bibfnamefont{T.}~\bibnamefont{Giamarchi}},
  \bibnamefont{and} \bibinfo{author}{\bibfnamefont{P.~L.}
  \bibnamefont{Doussal}}, \bibinfo{journal}{Phys. Rev. B}
  \textbf{\bibinfo{volume}{65}} (\bibinfo{year}{2002}).

\bibitem[{\citenamefont{Ewald}(1921)}]{ewald}
\bibinfo{author}{\bibfnamefont{P.~P.} \bibnamefont{Ewald}},
  \bibinfo{journal}{Ann. Phys.} \textbf{\bibinfo{volume}{64}},
  \bibinfo{pages}{253} (\bibinfo{year}{1921}).

\bibitem[{\citenamefont{Abramowitz and Stegun}(1970)}]{stegun}
\bibinfo{author}{\bibfnamefont{M.}~\bibnamefont{Abramowitz}} \bibnamefont{and}
  \bibinfo{author}{\bibfnamefont{I.~A.} \bibnamefont{Stegun}},
  \emph{\bibinfo{title}{Handbook of Mathematical Functions}}
  (\bibinfo{publisher}{Dover Publications}, \bibinfo{address}{New York},
  \bibinfo{year}{1970}).

\bibitem[{\citenamefont{Gold and Ghazali}(1990)}]{gold}
\bibinfo{author}{\bibfnamefont{A.}~\bibnamefont{Gold}} \bibnamefont{and}
  \bibinfo{author}{\bibfnamefont{A.}~\bibnamefont{Ghazali}},
  \bibinfo{journal}{Phys. Rev. B} \textbf{\bibinfo{volume}{41}},
  \bibinfo{pages}{7626} (\bibinfo{year}{1990}).

\bibitem[{\citenamefont{Schulz}(1993)}]{schulzwigner}
\bibinfo{author}{\bibfnamefont{H.~J.} \bibnamefont{Schulz}},
  \bibinfo{journal}{Phys. Rev. Lett.} \textbf{\bibinfo{volume}{71}},
  \bibinfo{pages}{1864} (\bibinfo{year}{1993}).

\bibitem[{\citenamefont{Bonsall and Maradudin}(1977)}]{bonsall}
\bibinfo{author}{\bibfnamefont{L.}~\bibnamefont{Bonsall}} \bibnamefont{and}
  \bibinfo{author}{\bibfnamefont{A.~A.} \bibnamefont{Maradudin}},
  \bibinfo{journal}{Phys. Rev. B} \textbf{\bibinfo{volume}{15}},
  \bibinfo{pages}{1959} (\bibinfo{year}{1977}).

\bibitem[{\citenamefont{Crandall}(1973)}]{crandall}
\bibinfo{author}{\bibfnamefont{R.~S.} \bibnamefont{Crandall}},
  \bibinfo{journal}{Phys. Rev. A} \textbf{\bibinfo{volume}{8}},
  \bibinfo{pages}{2136} (\bibinfo{year}{1973}).

\bibitem[{\citenamefont{Glasson et~al.}(2001)\citenamefont{Glasson, Dotsenko,
  Fozooni, Lea, Bailey, Papageorgiou, Andresen, and Kristensen}}]{wire}
\bibinfo{author}{\bibfnamefont{P.}~\bibnamefont{Glasson}},
  \bibinfo{author}{\bibfnamefont{V.}~\bibnamefont{Dotsenko}},
  \bibinfo{author}{\bibfnamefont{P.}~\bibnamefont{Fozooni}},
  \bibinfo{author}{\bibfnamefont{M.~J.} \bibnamefont{Lea}},
  \bibinfo{author}{\bibfnamefont{W.}~\bibnamefont{Bailey}},
  \bibinfo{author}{\bibfnamefont{G.}~\bibnamefont{Papageorgiou}},
  \bibinfo{author}{\bibfnamefont{S.~E.} \bibnamefont{Andresen}},
  \bibnamefont{and}
  \bibinfo{author}{\bibfnamefont{A.}~\bibnamefont{Kristensen}},
  \bibinfo{journal}{Phys. Rev. Lett.} \textbf{\bibinfo{volume}{87}},
  \bibinfo{pages}{176802} (\bibinfo{year}{2001}).

\bibitem[{\citenamefont{Shklovskii}(1999)}]{boris}
\bibinfo{author}{\bibfnamefont{B.~I.} \bibnamefont{Shklovskii}},
  \bibinfo{journal}{Phys. Rev. Lett.} \textbf{\bibinfo{volume}{82}},
  \bibinfo{pages}{3268} (\bibinfo{year}{1999}).

\end{thebibliography}
\end{document}